\documentclass[aps,prd,reprint,superscriptaddress]{revtex4-1}
\usepackage{hyperref}
\usepackage{graphicx}
\usepackage{enumerate}
\hypersetup{
    pdfnewwindow=true,      
    colorlinks=true,       
    linkcolor=blue,          
    citecolor=blue,        
    filecolor=blue,      
    urlcolor=blue           
}

\usepackage{subfigure}
\usepackage{amsmath}
\usepackage{amssymb}
\usepackage{amsfonts}
\usepackage{color}

\def\be{\begin{equation}}
\def\ee{\end{equation}}

\begin{document}

\title{Multimessenger Science Reach and Analysis Method for Common Sources of Gravitational Waves and High-energy Neutrinos}

\author{Bruny Baret}
\affiliation{AstroParticule et Cosmologie (APC), CNRS: UMR7164-IN2P3-Observatoire de Paris-Universit\'e Denis Diderot-Paris VII-CEA: DSM/IRFU, France}
\author{Imre Bartos}
\email[Corresponding author: ]{ibartos@phys.columbia.edu}
\affiliation{Department of Physics, Columbia University, New York, NY 10027, USA}
\author{Boutayeb Bouhou}
\affiliation{AstroParticule et Cosmologie (APC), CNRS: UMR7164-IN2P3-Observatoire de Paris-Universit\'e Denis Diderot-Paris VII-CEA: DSM/IRFU, France}
\author{Eric Chassande-Mottin}
\affiliation{AstroParticule et Cosmologie (APC), CNRS: UMR7164-IN2P3-Observatoire de Paris-Universit\'e Denis Diderot-Paris VII-CEA: DSM/IRFU, France}
\author{Alessandra Corsi}
\affiliation{LIGO Laboratory, California Institute of Technology, Pasadena, CA 91125, USA}
\author{Irene Di Palma}
\affiliation{Albert-Einstein-Institut, Max-Planck-Institut f\"ur Gravitationsphysik, D-30167 Hannover, Germany}
\author{Corinne Donzaud}
\affiliation{AstroParticule et Cosmologie (APC), CNRS: UMR7164-IN2P3-Observatoire de Paris-Universit\'e Denis Diderot-Paris VII-CEA: DSM/IRFU, France}
\affiliation{Universit\'e  Paris-sud, Orsay, F-91405, France}
\author{Marco Drago}
\affiliation{INFN, Gruppo Collegato di Trento and Universit\`{a} di Trento, I-38050 Povo, Trento, Italy}
\author{Chad Finley}
\affiliation{Oskar Klein Centre \& Dept. of Physics, Stockholm University, SE-10691 Stockholm, Sweden}
\author{Gareth Jones}
\affiliation{Cardiff University, Cardiff CF24 3AA, UK}
\author{Sergey Klimenko}
\affiliation{University of Florida, Gainesville, FL 32611, USA}
\author{Antoine Kouchner}
\affiliation{AstroParticule et Cosmologie (APC), CNRS: UMR7164-IN2P3-Observatoire de Paris-Universit\'e Denis Diderot-Paris VII-CEA: DSM/IRFU, France}
\author{Szabolcs M\'arka}
\affiliation{Department of Physics, Columbia University, New York, NY 10027, USA}
\author{Zsuzsa M\'arka}
\affiliation{Department of Physics, Columbia University, New York, NY 10027, USA}
\author{Luciano Moscoso}
\email[Deceased]{}
\affiliation{AstroParticule et Cosmologie (APC), CNRS: UMR7164-IN2P3-Observatoire de Paris-Universit\'e Denis Diderot-Paris VII-CEA: DSM/IRFU, France}
\author{Maria Alessandra Papa}
\affiliation{Albert-Einstein-Institut, Max-Planck-Institut f\"ur Gravitationsphysik, D-30167 Hannover, Germany}
\author{Thierry Pradier}
\affiliation{University of Strasbourg \& Institut Pluridisciplinaire Hubert Curien, Strasbourg, France}
\author{Giovanni Prodi}
\affiliation{INFN, Gruppo Collegato di Trento and Universit\`{a} di Trento, I-38050 Povo, Trento, Italy}
\author{Peter Raffai}
\affiliation{Department of Physics, Columbia University, New York, NY 10027, USA}
\affiliation{E\"otv\"os University, Institute of Physics, 1117 Budapest, Hungary}
\author{Virginia Re}
\affiliation{INFN, sezione di Roma Tor Vergata, I-00133 Roma Italy}
\author{Jameson Rollins}
\affiliation{LIGO Laboratory, California Institute of Technology, Pasadena, CA 91125, USA}
\author{Francesco Salemi}
\affiliation{Albert-Einstein-Institut, Max-Planck-Institut f\"ur Gravitationsphysik, D-30167 Hannover, Germany}
\affiliation{Leibniz Universit\"{a}t Hannover, D-30167 Hannover, Germany}
\author{Patrick Sutton}
\affiliation{Cardiff University, Cardiff CF24 3AA, UK}
\author{Maggie Tse}
\affiliation{Department of Physics, Columbia University, New York, NY 10027, USA}
\author{V\'eronique Van Elewyck}
\affiliation{AstroParticule et Cosmologie (APC), CNRS: UMR7164-IN2P3-Observatoire de Paris-Universit\'e Denis Diderot-Paris VII-CEA: DSM/IRFU, France}
\author{Gabriele Vedovato}
\affiliation{INFN, Sezione di Padova, I-35131 Padova, Italy}

\begin{abstract}
We present the baseline multimessenger analysis method for the joint observations of gravitational waves (GW) and high-energy neutrinos (HEN), together with a detailed analysis of the expected science reach of the joint search. The analysis method combines data from GW and HEN detectors, and uses the blue-luminosity-weighted distribution of galaxies. We derive expected GW+HEN source rate upper limits for a wide range of source parameters covering several emission models. Using published sensitivities of externally triggered searches, we derive joint upper limit estimates both for the ongoing analysis with the initial LIGO-Virgo GW detectors with the partial IceCube detector (22 strings) HEN detector and for projected results to advanced LIGO-Virgo detectors with the completed IceCube (86 strings). We discuss the constraints these upper limits impose on some existing GW+HEN emission models.
\end{abstract}

\maketitle

\section{Introduction}
\label{section:Introdution}

The observation of gravitational-waves (GWs) and neutrinos is entering a new and promising era with newly constructed detectors. GW observatories such as LIGO \cite{LIGO0034-4885-72-7-076901}, Virgo \cite{0264-9381-28-11-114002} and GEO \cite{2010CQGra..27h4003G} will be upgraded to second generation detectors within the next few years. Another advanced GW detector, LCGT \cite{LCGT0264-9381-27-8-084004}, is being constructed in Japan, while LIGO is considering plans for a third observatory in India \cite{LIGOindia}.

Both the emission mechanism and detection method of neutrinos can depend greatly on their energy, therefore it is worthwhile to consider different neutrino subgroups.  So far, only astrophysical MeV neutrinos (in addition to those from the Sun) have been detected, and only in one case, from supernova SN 1987A \cite{PhysRevLett.58.1490,PhysRevLett.58.1494}. Here we concentrate specifically on high-energy neutrinos (HENs): neutrinos $\gtrsim 100$ GeV that carry information about the particle acceleration region of astrophysical sources \cite{2008PhR...458..173B}.

HEN observatories currently in operation are IceCube \cite{IceCubeAhrens2004507}, a cubic-kilometer detector at the geographic South Pole, and \textsc{Antares} \cite{ANTARES} in the Mediterranean sea. \textsc{Antares} is being upgraded to a cubic-kilometer detector called KM3NeT in the following years \cite{deJong2010445}. A third HEN detector operating at the lake Baikal is also planned to be upgraded to a km$^3$ volume \cite{Avrorin2011S13}.

The joint (multimessenger) analysis of GW and HEN observations presents multiple advantages over single messenger analyses. Since both GWs and HENs are weakly interacting, their detection requires exceptionally high sensitivity. Search sensitivity can be greatly increased by combining data from GW and HEN detectors. The multimessenger approach could also significantly add to our understanding of the underlying mechanisms that create the astrophysical sources emitting both signals \cite{ET}. Aso \emph{et al.} \cite{asoicecube0264-9381-25-11-114039} designed a GW+HEN multimessenger search algorithm that uses the two LIGO detectors and IceCube to look for spatially and temporally coincident events. After requiring temporal coincidence between the GW and HEN signals within a given time window, the method reconstructs a ring on the sky based on the measured time delay between the signal arrival times in the LIGO detectors, and requires the direction of the neutrino signal to overlap with this ring. Aso \emph{et al.} show that such coincidence is extremely unlikely to arise from the background, making the requirement of coincidence very effective in reducing the false alarm rate of the joint measurement.

Pradier \cite{microquasarHENPradier2009268} considered a joint search with initial Virgo and \textsc{Antares}, and discussed the feasibility of a time coincident search using these two detectors. Pradier discussed microquasars and flares from soft gamma repeaters as plausible galactic GW+HEN sources. He also considered the detectability of quantum gravity effects by measuring the time delay between the arrival of GW and HEN signals.

We are close to the milestone of finishing the first coincident search for GWs and HENs for the initial LIGO-Virgo (S5/VSR1 science runs) and the partial \textsc{Antares} detector in its 5-string configuration. The analysis, a simpler version of what is presented below, uses the directional distribution and the time of arrival of HENs to trigger a GW follow-up analysis, similar to the analysis used for GW follow-up searches of GRBs (e.g. \cite{0004-637X-715-2-1438}). There are $\sim$200 neutrino triggers from the \textsc{Antares}, most of which are detected by digital optical modules (DOMs) two strings, while 13 neutrinos are detected by DOMs on three strings. The first scientific results of this search will be published soon.

In this article we introduce a joint GW and HEN analysis algorithm that combines the observations of a network of GW detectors and a HEN detector. Besides looking for astrophysical GW+HEN messengers, the search algorithm can also be used to derive upper limits on the population and flux of GW+HEN sources. We estimate the anticipated science reach for initial and advanced detectors, and discuss some of the existing emission models and how they can be constrained in the event of non-detection.

The distribution of astrophysical GW+HEN sources at detectable distances is not uniform. This can be used in a joint search algorithm to reduce false coincidences and increase sensitivity. One can weigh event candidates based on the expected source density in their direction. The method utilizes the blue-luminosity-weighted galaxy distribution to favor astrophysical sources over background coincidences. Blue luminosity is a good tracer of the rate of star formation and therefore CCSN rate (e.g. \cite{1999A&A...351..459C}). Long GRB rates in galaxies also correlate with the galaxies' blue luminosity \cite{2010MNRAS.405...57S} (typically long GRBs are more likely to occur in smaller, bluer galaxies compared to CCSNe. The method weighs source directions with the expected detectable source rate from the given directions, assuming that source distribution follows blue-luminosity distribution. We take the blue-luminosity distribution of galaxies up to 40~Mpc from the Gravitational Wave Galaxy Catalogue (GWGC) \cite{0264-9381-28-8-085016}.

The article is organized as follows. In section \ref{section:sources} we discuss some anticipated astrophysical sources of GWs and HENs. Section \ref{section:sciencereach} discusses the expected science reach of joint GW+HEN searches by presenting the interpretation of expected results for constraining existing emission and population models. Section \ref{section:gwhendata} describes GW and HEN detectors and data. In section \ref{section:jointanalysis} we introduce the baseline joint GW+HEN analysis, also discussing its relation to the presented science reach. Section \ref{section:conclusion} presents a summary of the method and the science reach.

\section{Astrophysical Sources}
\label{section:sources}

GWs and HENs may originate from a number of common sources. Plausible sources include gamma-ray bursts (GRBs) \cite{waxmanbachall,UHEN,HENfromSuccessfulnChokedGRB,2003PhRvL..90t1103G,2003PhRvL..90x1103R,precursorneutrinos,2003PhRvL..91g1102D,2006PhRvL..97e1101M,2006ApJ...651L...5M}, core-collapse supernovae (CCSNe), soft gamma repeaters \cite{Ioka:2001,2005ApJ...633.1013I,2006PhRvL..97v1101A} and microquasars \cite{ET,2009IJMPD..18.1655V}. For a joint GW+HEN search, potentially the most interesting sources are those which are difficult to detect using electromagnetic (EM) telescopes. Below we concentrate on GRBs as plausible GW+HEN sources. We discuss the scenarios in which GRBs have limited EM emission.

GRBs are exceptionally luminous gamma-ray flashes of cosmic origin \cite{ET}. They are thought to originate from at least two distinct types of astrophysical sources. The core-collapse of massive stars is thought to be the progenitor of at least some long-soft GRBs \cite{supernovaGRB2006ARA&A..44..507W}, while short-hard GRBs are usually associated with the mergers of compact binaries, such as two neutron stars (NS), or NS - black hole (BH) binaries \cite{Nakar06}.

Both types of GRB progenitors are expected to emit GWs. In the collapsar scenario, transient GW bursts are likely to be emitted by a variety of processes inside the star. These processes include rotating core collapse and bounce, non-axisymmetric rotational instabilities, postbounce convective overturn, or non-radial PNS pulsations \cite{GWsignaturefromCoreCollapse}. Other mechanisms, such as the fragmentation of collapsar accretion disks \cite{1538-4357-579-2-L63,1538-4357-630-2-L113,0004-637X-658-2-1173}, or suspended accretion \cite{PhysRevD.69.044007}, might also play an important role in GW emission.

Compact binary mergers are expected to be strong GW emitters in the sensitive frequency band of Earth-based GW detectors \cite{0004-637X-715-2-1453}. Binary systems lose angular momentum due to the emission of GWs. In the inspiral phase, the distance between the two compact objects decreases, while the rotational frequency increases. The process is expected to continue until the two objects merge, an event which is anticipated to involve an intermediate stage of a central quasi-axi-symmetric object surrounded by an accretion disk \cite{Nakar06}, producing a strong GW transient \cite{0004-637X-715-2-1453}.

Predictions on the energy radiated away through GWs from GRB progenitors vary over several orders of magnitude, depending on the emission model considered. Numerical simulations of various non-rotating CCSN progenitor models give GW emission of $10^{-8}-10^{-4}$ M$_\odot$c$^2$ \cite{GWsignaturefromCoreCollapse,0004-637X-655-1-416}, at typical frequencies of $500-700$ Hz.
Recent simulations of rotating CCSN progenitors with initial conditions chosen to resemble likely long-GRB progenitors show GW emission of $E_{\textsc{gw}} \sim 10^{-7}$~M$_{\odot}$c$^2$ \cite{2011PhRvL.106p1103O}, with characteristic frequencies of $f_c\sim500-1000$~Hz. Such emissions are detectable with advanced GW detectors from sources within our Galaxy.
Other, analytical and numerical models of long-GRBs that consider GW production by matter accretion around a central black hole tend to predict significantly stronger emission of GWs. For instance the fragmentation of collapsar accretion disks \cite{1538-4357-579-2-L63,1538-4357-630-2-L113,0004-637X-658-2-1173} could emit $10^{-3}-10^{-2}$ M$_\odot$c$^2$ in GWs, potentially in the most sensitive frequency band of LIGO-Virgo ($\sim150$ Hz). Recent simulations of black hole-torus systems show that non-axisymmetric instabilities in such systems may produce strong, quasi-periodic GW signals in the sensitive frequency band of ground based detectors, potentially detectable with advanced detectors from up to $\sim100$~Mpc \cite{2001PhRvL..87i1101V,PhysRevLett.106.251102}.
According to the suspended accretion model \cite{PhysRevD.69.044007}, GW emission up to $10^{-2}-10^{-1}$ M$_\odot$c$^2$ is possible in the most sensitive frequency band. Models for the likely progenitors of short GRBs, i.e. black hole - neutron star (NS) or NS-NS binaries, are expected to emit up to $10^{-1}$ M$_\odot$c$^2$ in GWs in the most sensitive frequency band \cite{2002ASPC..263..333J}.

HENs ($\gtrsim 100$ GeV neutrinos) are thought to be produced by various sources, including GRBs, as described by the internal shock model (e.g.~\cite{waxmanbachall,UHEN,HENfromSuccessfulnChokedGRB,precursorneutrinos}). According to the model, a central engine accelerates protons and electrons to relativistic velocities through Fermi acceleration~\cite{waxmanbachall,UHEN2,Waxman00}. Relativistic electrons emit gamma-rays through synchrotron radiation, while relativistic protons interact with these gamma-rays ($\gamma p$) or with other protons, ($pp$) producing charged pions ($\pi^{\pm}$) and kaons ($K^{\pm}$). Charged pions and kaons create HENs through the decay process \cite{HENAndoPhysRevLett.95.061103}
\begin{equation}
\pi^{\pm}, K^{\pm} \rightarrow \mu^{\pm} + \nu_{\mu}(\overline{\nu}_{\mu})
\label{piondecay}
\end{equation}
Muons may further decay and emit an additional muon-neutrino and electron-neutrino. However, muons may interact or undergo radiative cooling before they decay, in which case the contribution of secondary high-energy neutrinos will be reduced \cite{HENAndoPhysRevLett.95.061103,HeneventratePhysRevLett.93.181101}.

To characterize the HEN flux of astrophysical sources, we use the average number of detected HENs from a HEN source at 10 Mpc in a random direction, denoted by $n_{\textsc{hen}}$. This number depends on the sensitivity of the neutrino detector.

The Waxman-Bahcall model \cite{waxmanbachall}, the benchmark model of HEN emission from GRBs, predicts about $n_{\textsc{hen}} \approx 100$ neutrinos detected in a km$^3$ detector for a typical GRB at 10 Mpc (e.g. \cite{NeutrinoBATSEGuetta2004429}). Another model of HEN emission from mildly relativistic jets of core-collapse supernovae, and potentially from choked GRBs (below), predicts HEN emission of $n_{\textsc{hen}} \approx 10$ \cite{HENAndoPhysRevLett.95.061103} (note that the result presented in \cite{HENAndoPhysRevLett.95.061103} is three times higher, as it does not take into account neutrino flavor mixing). Horiuchi and Ando \cite{chokedfromreverseshockPhysRevD.77.063007} estimate $n_{\textsc{hen}}$ from reverse shocks in mildly relativistic jets to be $n_{\textsc{hen}} \approx 0.7-7$ for a km$^3$ neutrino detector (after taking into account neutrino flavor mixing). M{\'e}sz{\'a}ros and Waxman \cite{HENfromSuccessfulnChokedGRB} predict a emission of $10^{-1}$ - 10 detected neutrinos by a km$^3$ neutrino detector at a cosmological distance of $z\sim1$ for collapsars. Razzaque \emph{et al.} \cite{HeneventratePhysRevLett.93.181101} obtain $n_{\textsc{hen}}\approx 0.15$ for supernovae with mildly relativistic jets with jet energy of $E\sim10^{51.5}$ erg. Razzaque \emph{et al.} find this result to scale linearly with jet energy (i.e. it is $\sim \times10$ higher for typical hypernovae).

The joint search for astrophysical GW+HEN sources is of special importance for sources which are barely or not at all observable through other messengers. Below we discuss some scenarios in which EM emission from GRBs is limited. The discovery of such sources is one of the most valuable scientific goals of joint GW+HEN searches.

\subsection{Choked GRBs}

Core-collapse supernovae (CCSN) can have observable gamma-ray emission only if the relativistic outflow from the central engine, that is responsible for the production of gamma-rays, breaks out of the star \cite{chokedfromreverseshockPhysRevD.77.063007}. The outflow can advance only as long as it is driven by the central engine. If the duration of the activity of the central engine is shorter than the breakout time of the outflow, the outflow is \emph{choked}, resulting in a choked GRB \citep{ChokedGRBPhysRevLett.87.171102}.

HENs, due to their weak interaction with matter, can escape from inside the stellar envelope, from depths gamma-rays cannot. Consequently choked GRBs, similarly to ``successful'' GRBs, may be significant sources of HENs \cite{imre1,chokedfromreverseshockPhysRevD.77.063007}. Choked GRBs are expected to emit GWs similarly to successful GRBs.

\subsection{Low-luminosity GRBs}

Low-luminosity (LL) GRBs~\cite{Galama:1998ea,1998Natur.395..663K,2004ApJ...609L...5M,2004Natur.430..648S,2006ApJ...645L.113C,2006Natur.442.1011P,2006Natur.442.1014S,MNL2:MNL20050} form a sub-class of long GRBs with sub-energetic gamma-ray emission. Of the six detected long-GRBs with spectroscopically confirmed SN associations, four are of the LL-GRB variety \cite{2011arXiv1107.1346B}. Although in general both of these classes of GRBs have been associated with luminous Type Ic SNe, the less energetic SN that accompanied low-luminous burst GRB 060218 suggests that the connection may extend towards lower energy SN explosions (e.g. \cite{2006ApJ...645L.113C}). In addition, the relative close proximity of observed LL-GRBs implies a much higher rate of occurrence than that of canonical high-luminosity (HL) GRBs \cite{2006ApJ...645L.113C,2006Natur.442.1014S}. Due to this higher population rate, the total diffuse HEN flux from LL-GRBs can be comparable or even surpass the flux from conventional HL GRBs ~\cite{2006ApJ...651L...5M,2007APh....27..386G,2007PhRvD..76h3009W}. While individual LL GRBs are less luminous in neutrinos, the higher population rate makes LL-GRBs valuable sources for GW+HEN searches.

\subsection{Unknown Sources and Mechanisms}

A potential advantage of the search for astrophysical GW and HEN signals is the discovery of previously unanticipated sources or mechanisms. Such mechanisms include, for example, HEN emission at a larger beaming angle than the observed beaming angle for gamma rays.

\subsection{Source population}

Some of the most interesting GW+HEN sources, CCSNe (SNe Ib/c) are expected to be similarly or somewhat less abundant \cite{HENAndoPhysRevLett.95.061103} than galactic SNe, with an estimated rate of $\sim$1/century/Milky Way equivalent (MWE) galaxy \cite{2011MNRAS.412.1473L}. Additionally, late-time radio observations of supernovae indicate that $\lesssim1\%$ of SNe have mildly relativistic jets \cite{2010Natur.463..513S}. The expected rate of another interesting population, LL-GRBs, is $\sim 3\times10^{-5}$ yr$^{-1}$ / galaxy ($\sim$300 Gpc$^{-3}$yr$^{-1}$) \cite{Liang2007}, although this value is highly uncertain \cite{2011MNRAS.410.2123H}. The rate of HL-GRBs is several orders of magnitudes smaller ($\sim1$ Gpc$^{-3}$yr$^{-1}$) \cite{Liang2007}.

\subsection{Time Delay between Gravitational-waves and High-energy Neutrinos}

GWs and HENs are though to be emitted through different emission mechanisms by a joint source. Due to this difference the two messengers will likely be observed with a time delay. Further time delay can arise from the randomness of HEN-detection throughout the emission period. The time difference between the observation of GWs and HENs can be an important factor in the interpretation of multimessenger signals \cite{Baret20111}. For instance an HEN detected prior to a GW signal from a GRB would indicate precursor activity in the GRB, while the time delay of the earliest HEN after the GW signature of a collapsar may be indicative of jet propagation within the stellar envelope. Besides interpretation, an upper bound on the temporal difference between the observation of GWs and HENs is an important parameter in designing a joint search algorithm (see Section \ref{section:CoincidenceWindow} below).

Baret et al. \cite{Baret20111} recently published an estimate on the upper bound of the time delay between GWs and HENs from GRBs. Their analysis focused on GRBs as arguably the most promising multimessenger GW+HEN sources. They obtained a conservative $\sim500$~s time window that took into account processes motivated by current GRB models. The duration of each process was determined based on electromagnetic observational data.

\section{Science Reach}
\label{section:sciencereach}

In this section we investigate the constraints one can introduce with the GW+HEN search on the population of astrophysical GW+HEN sources. Below we first estimate the expected population upper limits from the GW+HEN search as a function of source parameters, after which we interpret these constraints. The science reach analysis presented here follows the method of Bartos \emph{et al.} \cite{PhysRevLett.107.251101}, that we outline below.

\begin{figure*}
\begin{center}
\resizebox{1\textwidth}{!}{\includegraphics{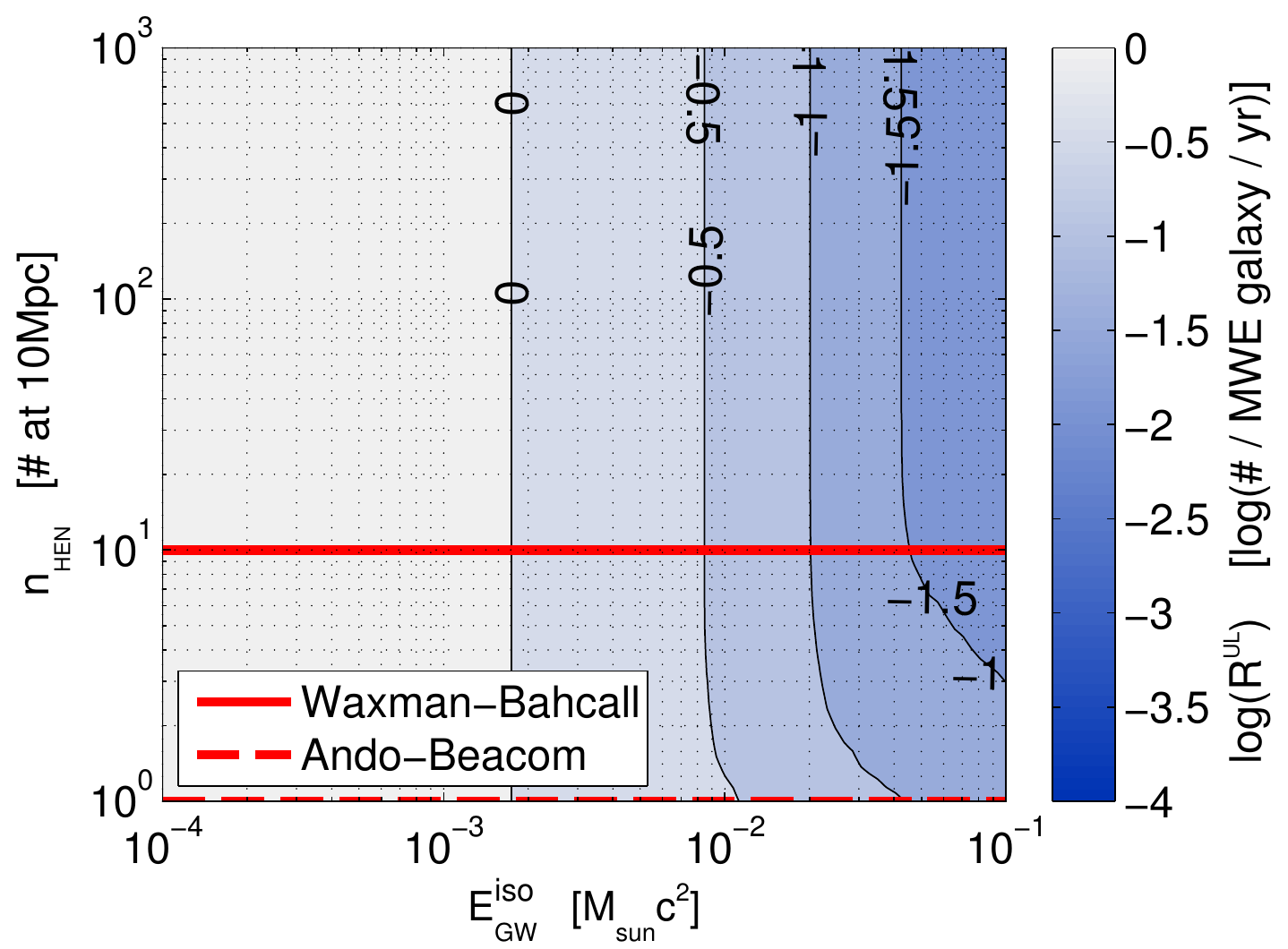}\includegraphics{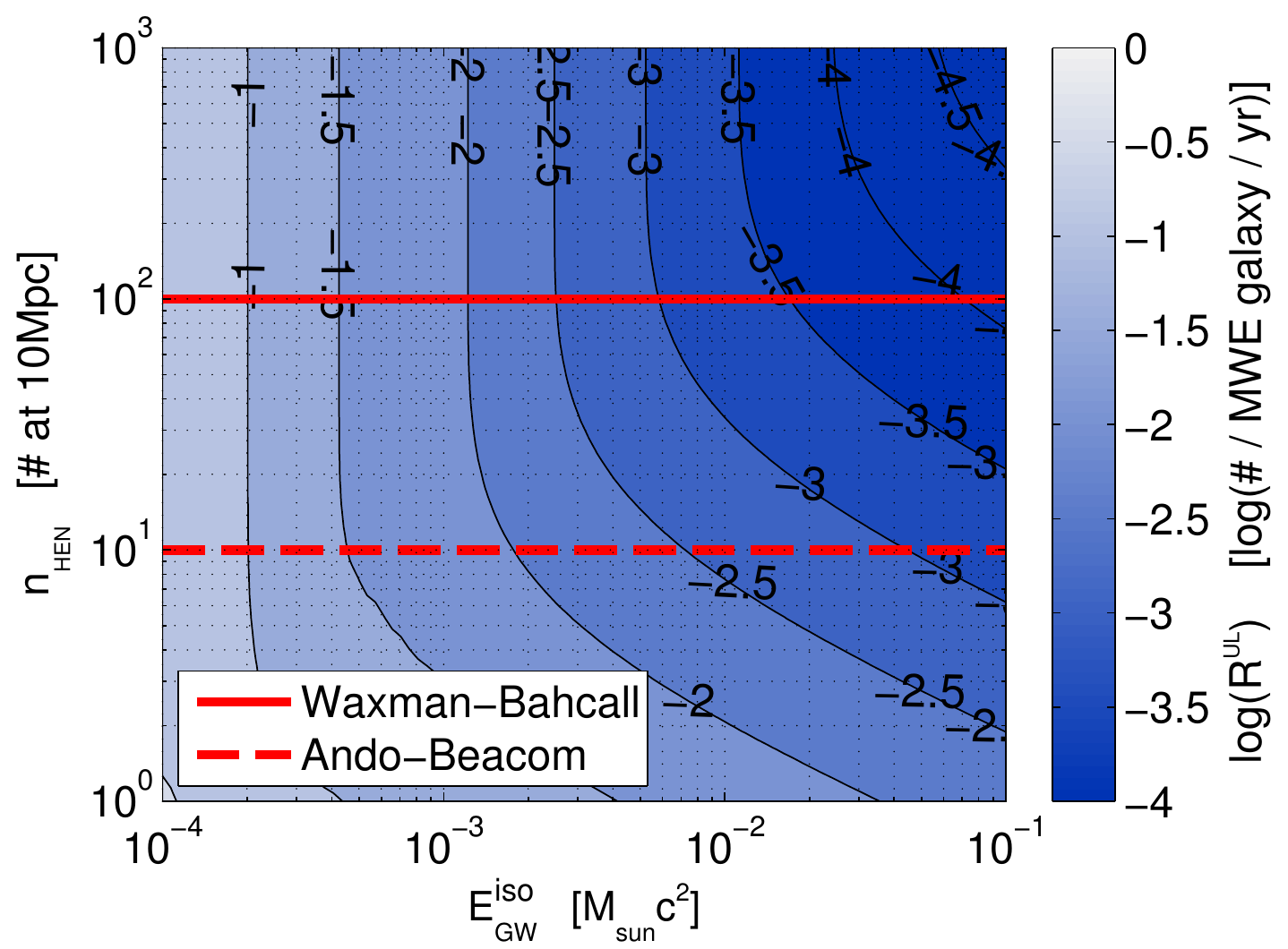}}
\end{center}
\caption{
Expected GW+HEN source population upper limits for IceCube-22 coincident with initial LIGO-Virgo (left) and IceCube-86 coincident with advanced the LIGO-Virgo detectors (right; courtesy of \cite{PhysRevLett.107.251101}), with one year of coincident measurement time. The results take into account the blue-luminosity-weighted galaxy distribution. The x-axis represents the GW energy output of a standard source. The y-axis represents the number of detected neutrinos from a standard source at 10~Mpc. The color scale shows the obtained source rate upper limit $R^{UL}$ in logarithmic units of number of sources per (Milky Way equivalent) galaxy per year.  On both plots, the two horizontal lines (scaled for detector sensitivity) show the Waxman-Bahcall emission model \cite{NeutrinoBATSEGuetta2004429} (higher) and the HEN emission model of Ando and Beacom \cite{HENAndoPhysRevLett.95.061103} for reverse shocks in mildly relativistic supernova jets / choked GRBs (lower).} \label{figure:individualpopulationestimate}
\end{figure*}

In determining the GW+HEN population upper limit we assume standard GW+HEN sources with the same intrinsic emission.
Limits based on a fixed average brightness are conservative compared to those using any other brightness distribution.
We consider maximum one HEN detected for each source. This is a reasonable (and conservative) assumption given that there has been no discovery with neutrino detectors. We introduce the exclusion distance $D_{50\%}^{\textsc{gwhen}}$, which is the maximum distance that satisfies the following criterion: for an astrophysical GW+HEN 
burst at a distance $<D_{50\%}^{\textsc{gwhen}}$ in a 
typical direction and with one detected HEN, the probability that it is more significant (Eq. \ref{totalsignificance}) than the loudest observed event of the GW+HEN search is $\geq50\%$. This distance depends on the total (isotropic-equivalent) energy emitted in GWs ($E_{GW}^{iso}$) of a GW+HEN source. Using this distance, we calculate the minimum astrophysical GW+HEN source rate (i.e. population) that would have produced at least one detected astrophysical HEN signal with $\gtrsim 90\%$ probability. This source rate will be the rate upper limit.

To estimate the expected results with the GW+HEN search, we approximate the sensitivity of the GW+HEN algorithm with that of published externally triggered GW searches (e.g. \cite{0004-637X-715-2-1438,PhysRevD.76.062003}). This is a reasonable (and conservative) approximation if one chooses the threshold for GW and HEN trigger selection such that there will be $O(1)$ spatially and temporally coincident GW and HEN signals for the duration of the measurement. In an externally triggered search for GW bursts in coincidence with GRBs, Abbott \emph{et al.} \cite{0004-637X-715-2-1438} obtained a median upper bound of $h_{rss}^{ext}\approx3.8\times10^{-22}$ strain root sum square using the initial LIGO-Virgo GW detector network (S5/VSR1 science run). They used a sine-Gaussian waveform with characteristic frequency of $\sim150$ Hz, which is in the most sensitive frequency band of the GW detectors. This upper limit corresponds to the minimal GW signal amplitude that, with $90\%$ confidence, produces larger significance than the loudest joint GW+HEN event in the real data measured in coincidence with an external trigger. To estimate the performance of advanced detectors (advanced LIGO-Virgo), we estimate their median strain upper bound as 0.1 times that of initial detectors (i.e. $3.8\times10^{-23}$). We note here that with additional advanced detectors, such as LIGO-Australia \cite{LIGOAUSTRALIA} and LCGT \cite{LCGT0264-9381-27-8-084004}, the sensitivity of the GW detector network will further increase. For comparison, we note that the upper bound obtained with the all-sky GW search of Abadie \emph{et al.} \cite{PhysRevD.81.102001} for sine-Gaussian signals at $\sim150$~Hz with the initial LIGO-Virgo detector network is $h_{rss}^{all-sky}\approx6\times10^{-22}$. The all-sky search corresponds to the lower limit for the sensitivity of GW searches as no additional information is used besides GW data.

For the GW+HEN search we introduce the upper bound $h_{rss}^{\textsc{gwhen}}$, and estimate this upper bound to be $h_{rss}^{\textsc{gwhen}}\approx h_{rss}^{ext}$. We assume that a GW signal with $h_{rss} \geq h_{rss}^{\textsc{gwhen}}$ in coincidence with a detected astrophysical HEN would be detected by the joint GW+HEN search with $\geq 90\%$ probability. Given $h_{rss}^{\textsc{gwhen}}$ (i.e. the amplitude at the detector) and $E_{GW}^{iso}$ (i.e. the amplitude at the source), we can calculate the radius within which there was likely ($P\geq90\%$) no GW+HEN event from which a HEN was detected. This distance, averaged over all directions on the sky, is \cite{0004-637X-715-2-1438}
\begin{equation}
D^{\textsc{gwhen}} = 12 \mbox{ Mpc}\left(\frac{E_{GW}^{iso}}{10^{-2}\mbox{ M}_{\odot}\mbox{c}^2}\right)^{1/2} \left(\frac{h_{rss}^{ext}}{h_{rss}^{\textsc{gwhen}}}\right)
\end{equation}

From the fact that no astrophysical HEN was detected from a GW+HEN source within $D^{\textsc{gwhen}}$, we obtain a source rate upper limit as the highest source rate that would have produced at least one detected neutrino within $D^{\textsc{gwhen}}$ with $\lesssim90\%$ probability. Assuming a Poissonian source number distribution, this corresponds to an average source rate of 2.3 over one year long measurement. We denote this source rate upper limit by $R^{\textsc{ul}}$. To obtain $R^{\textsc{ul}}$, we calculate the average number of sources within $D^{\textsc{gwhen}}$ over the duration of the measurement from which at least one neutrino has been detected. $R^{\textsc{ul}}$ will be the rate that corresponds to 2.3 detected sources on average. The rate depends on the blue-luminosity-weighted galaxy distribution within $D^{\textsc{gwhen}}$ (see section \ref{section:sourcedist}), as well as the neutrino flux ($n_{\textsc{hen}}$) from a standard source.

To scale theoretical expectations on the HEN rate for km$^3$ detectors to expectations for the IceCube 22 string (hereafter IceCube-22) detector (which was operating the same time as the initial LIGO-Virgo detectors (S5/VSR1 science run)), we estimate IceCube-22 to be approximately 10 times less sensitive.

The estimated source rate upper limit is dependent on the beaming of HEN emission (beaming is less significant for GWs). The beaming of HEN emission is uncertain, but it is probably similar to the beaming of gamma rays, as the two emission mechanisms are connected. For this reason we use the gamma-ray beaming factor obtained for LL-GRBs, estimated to be less than 14 \cite{Liang2007}. The obtained upper limits scale linearly with the beaming factor (since we only expect to see sources for which the beam points towards us).

We calculate population upper limits for a range of GW isotropic emission $E_{GW}^{iso}$ and neutrino emission $n_{\textsc{hen}}$. The results are shown in Figure \ref{figure:individualpopulationestimate}, both for initial and advanced detectors.

\begin{figure}
\begin{center}
\resizebox{0.475\textwidth}{!}{\includegraphics{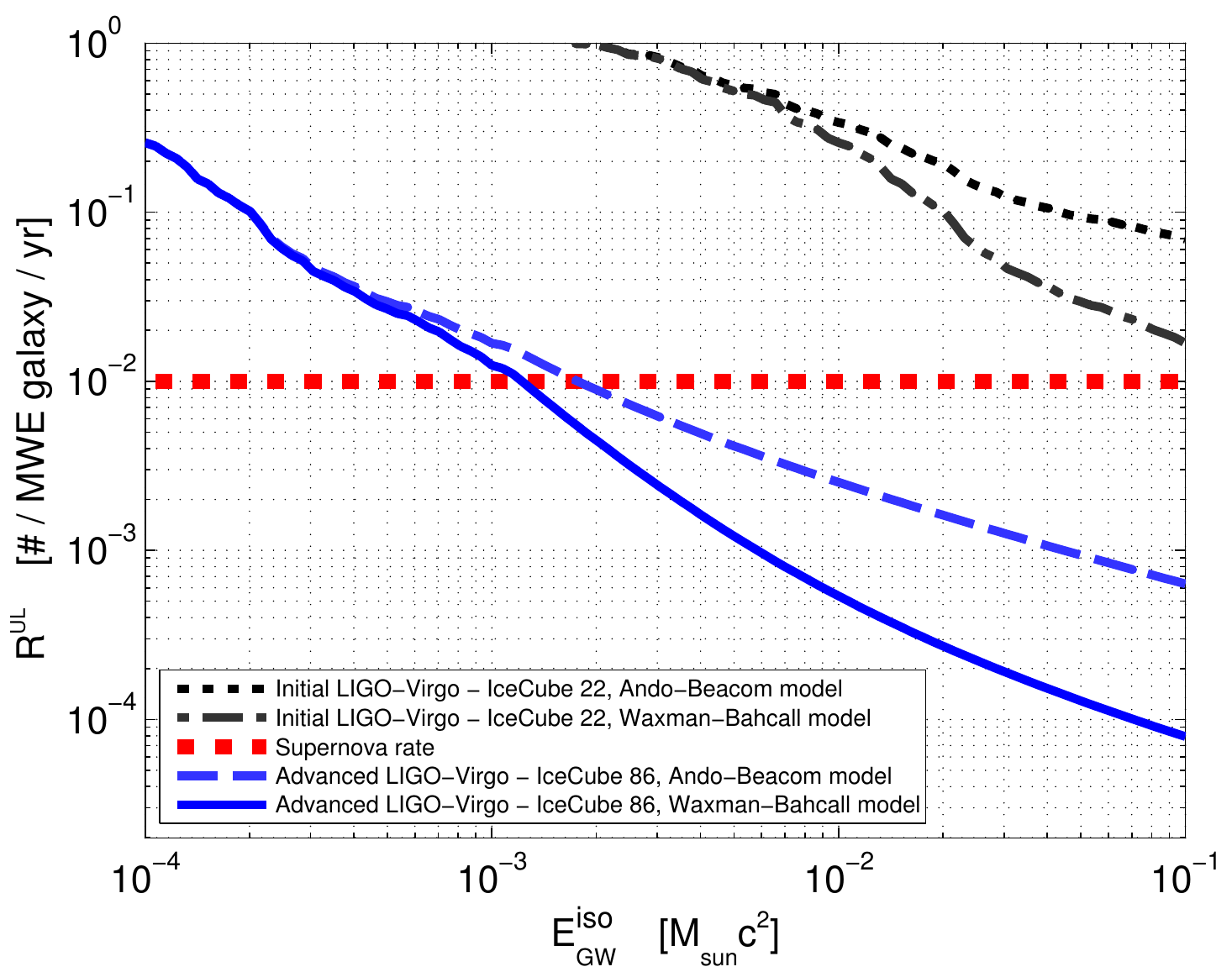}}
\end{center}
\caption{Expected GW+HEN source population upper limits for anticipated HEN emission from two emission models, as functions of isotropic-equivalent GW emission energy $E_{GW}^{iso}$. Results are shown both for measurements with the initial LIGO-Virgo detectors and the IceCube-22 detector (dashed line), as well as for the advanced LIGO-Virgo detectors and the IceCube-86 detector (solid line), both with one year of coincident measurement time. For comparison, the galactic supernova rate is shown (dotted line). This Figure shows a subset of the results shown in Figure \ref{figure:individualpopulationestimate}.} \label{figure:populationmildjets}
\end{figure}

To interpret the science reach of the expected GW+HEN population upper limits described above, we consider the source parameters from some of the emission models, as discussed in section \ref{section:sources}. In Figure \ref{figure:individualpopulationestimate} horizontal lines indicate the neutrino rate predictions of the Waxman-Bahcall emission model \cite{waxmanbachall}, as well as the emission model for mildly relativistic jets by Horiuchi and Ando \cite{chokedfromreverseshockPhysRevD.77.063007}. The population upper limit estimates for these two models specifically, as functions of $E_{GW}^{iso}$ down to $E_{\textsc{gw}}=10^{-4}$~M$_\odot$c$^2$, are shown in Figure \ref{figure:populationmildjets}. For sources of weaker GWs than $10^{-4}$~M$_\odot$c$^2$ as predicted by some CCSN simulations \cite{GWsignaturefromCoreCollapse,0004-637X-655-1-416,2011PhRvL.106p1103O}, observations will focus on galactic sources.

\section{Data}
\label{section:gwhendata}

In this section we describe the output of the GW and HEN detectors, as well as the astrophysical source distribution. We derive the quantities that will be used in joint GW+HEN analyses.

\subsection{Gravitational-wave Data}
\label{gw}

Interferometric GW detectors measure a passing GW by monitoring the distance between test masses using laser interferometry \cite{2003CQGra..20S.853C}. The detectors considered in the present analysis are Michelson interferometers, in which a laser beam is divided into two perpendicular laser beams directed along the two arms of the interferometer. The two arms, with roughly equal length $L$, contain resonant Fabry-Perot optical cavities with partially transmitting input mirrors and highly reflective end mirrors \cite{LIGO0034-4885-72-7-076901}. A passing GW changes the phase of the laser light between the entering and exiting beams with $180^\circ$ phase shift between the two perpendicular arms. The induced phase shift is measured through the interference of the outcoming beams from the two arms, and is used to calculate the so-called \emph{differential arm length} $\Delta L = L_1 - L_2$ with $L_1$ and $L_2$ being the lengths of the two arms. The quantity $h(t) = \Delta L / L$ is called the \emph{GW strain}, and is used to define the amplitude of a GW.

The measured GW strain $h(t)$ is defined by the passing GW as
\begin{equation}
h(t) = F_+h_+(t) + F_\times h_\times(t)
\label{hoft}
\end{equation}
where $+$ ("plus") and $\times$ ("cross") are the two polarizations of the GW, at $45^\circ$ angles from each other. The coefficients $F_+$ and $F_\times$ are the so-called antenna factors that depend on the direction of the incoming GW relative to the orientation of the detector, as well as the polarization of the GW (see, e.g. \cite{PhysRevD.63.042003}). $h_+(t)$ and $h_\times(t)$ are the amplitudes of the GW in the two polarizations.

GW search algorithms are designed to detect and extract information about a GW signal from a stream of strain data from a set of GW detectors (e.g. \cite{skymap0264-9381-26-15-155017,Xpipeline1367-2630-12-5-053034,cWB0264-9381-25-11-114029}). One can think of a generic search algorithm as a \emph{GW radiometer}, outputting the excess GW energy measured by a network of detectors, as a function of time $t$ and sky location $\overrightarrow{x_s}$. These data analysis algorithms usually output so-called \emph{GW triggers}, potential GW signals whose significance exceed a given threshold. A GW trigger's significance is characterized by a suitable test statistic (see e.g. \cite{skymap0264-9381-26-15-155017,Xpipeline1367-2630-12-5-053034,skymap0264-9381-26-15-155017}). GW triggers can have additional parameters, such as time of arrival, amplitude or waveform. Data from a network of GW detectors can also be used to recover directional information (e.g., \cite{positionreconstructionfairhurst1367-2630-11-12-123006}).

For the purposes of the joint GW+HEN analysis, we consider short transient events. The duration of a transient GW event is expected to be much shorter than the coincidence time window \cite{Baret20111} of GW and HEN events (the window in which all GW and HEN signals arrive).

To obtain the background distribution of GW triggers, we time-shift data from the different GW detectors compared to each other such that no astrophysical signal can appear in more than one detector at a time. Background triggers can be generated in such a way for many different time shifts.

Let us assume that we have a GW search algorithm that identifies a set of GW triggers, for each trigger calculating a test statistic TS and a skymap (point spread function) $\mathcal{F}_{\textsc{gw}}(\overrightarrow{x_s})$. The point spread function gives the probability distribution of source direction, given that the GW event is of astrophysical origin. To calculate the significance of a joint event, we need to take into account TS as well as $\mathcal{F}_{\textsc{gw}}(\overrightarrow{x_s})$. The background distribution of TS can be obtained from time-shifted data. The distribution of TS for the case of a signal present, however, is not available, as it would greatly depend on the properties and direction of the signal. Therefore, we take into account TS in the joint significance by calculating its p-value, given the background distribution.

Let FAR$_i$ be the false alarm rate of GW event $i$, corresponding to the rate of GW events with TS$\geq$TS$_i$ (average number of events over unit time). For TS$_i$ we assign a p-value of $$p_{\textsc{gw}}^{(i)}=1-\mbox{Pois}(0;T\cdot\mbox{FAR}_{i}),$$ where Pois$(k,\lambda)$ is the Poisson probability of $k$ outcome with $\lambda$ average, and $T$ is the coincidence time window (see section \ref{section:CoincidenceWindow}). FAR$_i$ is calculated from the distribution of background events.

For the skymap $\mathcal{F}_{\textsc{gw}}(\overrightarrow{x_s})$ both the background and signal distributions are available. We therefore take this information into account in the form of a likelihood ratio. Here our null hypothesis is that the GW event is a random fluctuation from the background, i.e. it has no preferred direction. We therefore approximate the background likelihood $\mathcal{B}_{\textsc{gw}}^{(i)}$ to be a flat distribution over the whole sky, i.e.
\begin{equation}
\mathcal{B}_{\textsc{gw}}^{(i)} = \frac{1}{4\pi}.
\end{equation}
Our alternative hypothesis is that GW event $i$ came from an astrophysical source at direction $\overrightarrow{x_s}$. The signal likelihood $\mathcal{S}_{\textsc{gw}}^{(i)}$ will be the calculated skymap, i.e.
\begin{equation}
\mathcal{S}_{\textsc{gw}}^{(i)}(\overrightarrow{x_s}) = \mathcal{F}_{\textsc{gw}}(\overrightarrow{x_s}).
\end{equation}

\subsection{High-energy Neutrino Data}
\label{nu}

HENs traveling through the Earth interact with the surrounding matter with a small interaction cross section.
In charged-current interactions, most of the neutrino's energy is carried away by a single high-energy electron, muon, or tau particle, which will emit Cherenkov radiation as it travels through the detector medium (water or ice).
For neutrino astronomy, the high-energy muons are generally the most useful: they neither lose energy as rapidly as electrons nor decay as rapidly as taus, and therefore can have paths many kilometers long.
The Cherenkov light emitted along this path can be detected and used to measure the direction and energy of the muon and infer similar information about the primary neutrino.
HEN detectors contain large numbers of optical sensors placed along vertical wires (strings). These optical sensors detect the Cherenkov photons emitted by muons. z

The direction of HENs can be reconstructed using the arrival time of Cherenkov photons at different optical sensors, with a precision of
$\sim 0.5^\circ-1^\circ$ (depending on energy) for IceCube \cite{2011arXiv1106.3484I},
or less than $0.3^\circ$ for \textsc{Antares} \cite{Antares2010arXiv1002.0754C}.
Direction reconstruction is also one of the major tools in background rejection.
So-called atmospheric muons,
 created by cosmic rays interacting with particles in the atmosphere over the detector, are the dominant background.
To suppress these events, searches for neutrinos are principally performed using up-going events, i.e. those that have traveled through Earth and therefore can be attributed only to a neutrino.
  The vast majority of these up-going neutrinos are themselves the result of cosmic ray interactions on the other side of the Earth.  These are the so-called atmospheric neutrinos, and in general constitute an irreducible background in searches for astrophysical neutrinos from space.

Many sources of astrophysical neutrinos are expected to exhibit a harder energy spectrum (typically $E^{-2}$) compared with the soft spectrum of atmospheric neutrinos ($\sim E^{-3.7}$).  In such cases, reconstructed energy can play a role in further separating signal from background.  The measured energy of the muon (related to the amount of light detected) becomes an estimated lower bound for the neutrino primary energy, since the muon may have propagated an unknown distance before reaching the detector.  Probability distribution functions for the muon energies expected from signal (of different source spectra) and background can then be used in likelihood analyses to enhance sensitivity for hard source spectra while retaining sensitivity to softer spectra
\cite{likelihood_ratio_icecube,2009ApJ...701L..47A}.

A given reconstructed neutrino event $i$ consists of a time of arrival $t_{\nu}^{(i)}$, reconstructed direction $\overrightarrow{x_i}$, directional uncertainty $\sigma_i$ and reconstructed neutrino energy $E_i$. The neutrino point spread function, i.e. the probability distribution of the neutrino source direction, is defined as
\begin{equation}
\mathcal{F}_{\nu}(\overrightarrow{x_s}|\overrightarrow{x_i}) = \frac{1}{2\pi\sigma^2}\cdot e^{-\frac{|\overrightarrow{x_i}-\overrightarrow{x_s}|^2}{2\sigma_i^2}},
\end{equation}
where $\overrightarrow{x_s}$ is the true sky location of the source. The HEN point spread function is incorporated in the joint GW+HEN significance in the form of a likelihood ratio. Our null hypothesis is that HEN event $i$ is a detected atmospheric neutrino, having no preferred direction. We approximate the background likelihood $\mathcal{B}_{\nu}^{(i)}$ to be a flat distribution over one hemisphere (since a neutrino detector is only sensitive to roughly half the sky), i.e.
\begin{equation}
\mathcal{B}_{\nu}^{(i)} = \frac{1}{2\pi}.
\end{equation}
Our alternative hypothesis is that the neutrino came from an astrophysical source at direction $\overrightarrow{x_s}$. The signal likelihood $\mathcal{S}_{\nu}^{(i)}$ will be the point spread function, i.e.
\begin{equation}
\mathcal{S}_{\nu}^{(i)}(\overrightarrow{x_s}) = \mathcal{F}_{\nu}(\overrightarrow{x_s}).
\end{equation}

For reconstructed neutrino energy $E_i$, the background distribution is known from the detected (background) neutrinos. The distribution of $E_i$ for astrophysical signals, however, depends on the source emission model, therefore we treat it as unknown. We therefore take into account $E_i$ in comparison with its background distribution by calculating its p-value $p_{\textsc{hen}}^{(i)}$, defined as the fraction of background neutrinos having energy $E\geq E_{i}$.

\subsection{Neutrino Clustering}
\label{section:NeutrinoClustering}

After obtaining the properties of individual neutrino events, we consider the possibility that multiple neutrinos are detected from the same astrophysical source. A set of neutrinos can potentially originate from the same source only if they are \emph{spatially} and \emph{temporally coincident}.

As it is described in section \ref{section:CoincidenceWindow} below, assuming that a GW+HEN source emits neutrinos within a time interval $\Delta t_{\nu}$ (denoted by $t_{\nu}^{(+)}-t_{\nu}^{(-)}$ in section \ref{section:CoincidenceWindow}), we consider a set of neutrinos to be temporally coincident if all neutrinos arrive within a $\Delta t_{\nu}$ time interval.

For spatial coincidence, we require that all neutrinos have a common direction of origin from where each neutrino can originate with probability above a threshold $P_{min}=0.05$ (i.e. if the probability that a neutrino came from a direction farther from its reconstructed direction than the common direction is $\leq P_{min}$). This probability threshold corresponds to an angular difference threshold $\Delta\overrightarrow{x}_{s,i}^{max}$ between the common direction and the neutrino direction ($\Delta\overrightarrow{x}_{s,i}^{max} = \sigma_i\sqrt{2\ln(1/P_{min})}$. The average angular distance threshold for IceCube-22 neutrinos is $\sim2.5^\circ$ for $P_{min}=0.05$). Neutrinos $i$ and $j$ will be spatially coincident if their angular distance is $\leq \Delta\overrightarrow{x}_{s,i}^{max} + \Delta\overrightarrow{x}_{s,j}^{max}$.

To take into account more than one neutrino in the joint GW+HEN significance, one needs to account for the probability of detecting a certain number of neutrinos from the background. The distribution of the number of neutrinos expected for astrophysical signals is model-dependent due to the non-homogenous source distribution, and is in general unknown. However, the probability of background neutrinos to occur in the same time window can be calculated. Given that we have at least one detected neutrino, the probability of detecting N neutrinos in the allowed time window is $\mbox{Pois}(N-1;f_{\nu}\Delta t_{\nu})$, where $f_{\nu}$ is the background neutrino rate (its typical value is $\sim10^{-4}$~Hz for IceCube-22 and $\sim10^{-2.5}$~Hz for the completed IceCube). We calculate the p-value for one neutrino to be in a time-window with N or more neutrinos to be
\begin{equation}
p_{cluster}(N) = 1 - \sum_{i=0}^{N-2} \mbox{Pois}(i;f_{\nu}\Delta t_{\nu}).
\end{equation}
This p-value is additionally taken into account to the other p-values in the joint test statistic. Note that, if one has only one detected neutrino in the cluster, $p_{cluster}(1)=1$.

Note that the decision of whether to treat coincident neutrinos as a cluster or as individual events is made by the analysis based on which combination yields the highest significance (this decision process can proceed iteratively on the remaining neutrinos until all are accounted for).

\subsection{Astrophysical Source Distribution - the Galaxy Catalog}
\label{section:sourcedist}

The distribution of astrophysical GW+HEN sources at detectable distances is not uniform. This can be used in a joint search algorithm to increase sensitivity. One can weigh event candidates based on the expected source density in their direction. The density of proposed GW+HEN sources can be connected to the blue luminosity of galaxies \cite{blueluminositybinary1991ApJ...380L..17P,2011arXiv1103.0695W}, while source density can also depend on, e.g., the galaxy type \cite{2006IJMPD..15..235D,2010ApJ...716..615O}. We take the blue-luminosity distribution of galaxies up to 40~Mpc from the Gravitational Wave Galaxy Catalogue (GWGC) \cite{0264-9381-28-8-085016}.

Let the astrophysical GW+HEN source density be $\rho(r,\overrightarrow{x})$, where $r$ and $\overrightarrow{x}=(\phi,\theta)$ are the source distance and direction on the sky, respectively. The probability distribution of astrophysical neutrinos as a function of direction is proportional to the number of sources in the given direction, weighted with the distance of the sources to the $-2^{nd}$ power (which cancels out in the volume integral):
\begin{equation}
\mathcal{F}_{gal}(\overrightarrow{x_s}) =  \frac{1}{N_{\nu}}\int_0^{D_{horizon}} \rho(r,\overrightarrow{x_s}) \mbox{d}r
\label{nugalaxydist}
\end{equation}
where $N_\nu$ is a normalization factor, $\overrightarrow{x_s}$ is the source direction and $D_{horizon}$ is an expectation-motivated cutoff distance
(see, e.g., \cite{galaxycatalog0004-637X-675-2-1459,0264-9381-28-8-085016,2010CQGra..27q3001A}).
For HEN searches, $D_{horizon}$ can be chosen to be very large, however for joint GW+HEN searches $D_{horizon}$ will be chosen to be the cutoff distance related to GW detection (see below).
For searches using initial GW detectors, a reasonable choice can be $D_{horizon} = 40$~Mpc. Given the detector sensitivities and typical source strengths, sources of interest father than this distance are unlikely to have measurable effect.
In the following we will refer to this distribution as the \emph{weighted galaxy distribution}.

\begin{figure}
\begin{center}
\resizebox{0.47\textwidth}{!}{\includegraphics{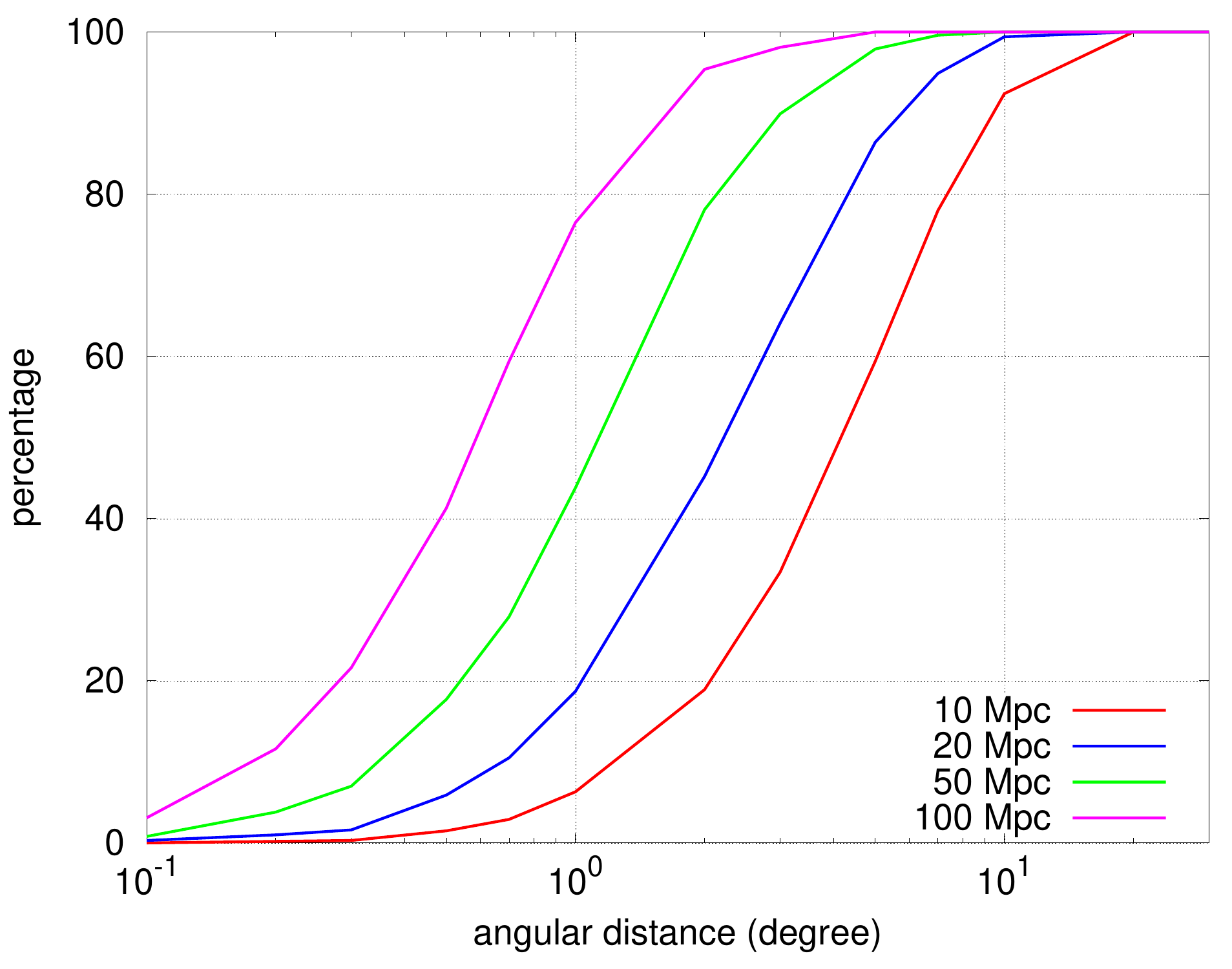}}
\end{center}
\caption{Probability that a random sky direction falls by chance within a given angular distance of at least one of the nearby galaxies listed in the GWGC.} \label{figure:galaxycatalogdemonstration}
\end{figure}

To take into account the galaxy distribution in the joint analysis, we consider our null hypothesis to be that the joint signal is a random coincidence from the background, i.e. it has no directional preference. This results in a background likelihood of
\begin{equation}
\mathcal{B}_{gal}^{(i)} = \frac{1}{2\pi},
\end{equation}
where we take into account that a joint event can only come from half of the sky due to the directional sensitivity of neutrino observatories. The alternative hypothesis is that the joint event came from an astrophysical source at direction $\overrightarrow{x_s}$. The corresponding signal likelihood is
\begin{equation}
\mathcal{S}_{gal}^{(i)}(\overrightarrow{x_s}) = \mathcal{F}_{gal}(\overrightarrow{x_s}).
\end{equation}

The information on the distribution of galaxies is accurate for directions outside the galactic plane. Within the plane, the large density of galactic stars makes it more difficult to detect galaxies in these directions. This incompleteness needs to be taken into account in our perceived source distribution. Another complication is that nearby galaxies
can be \emph{smeared} to a finite area of the sky. We ensure that no source is missed due to these incompletenesses by performing a complementary search with no galactic weighing. Such a search, while being significantly less sensitive than a search that takes into account the galaxy catalog, is capable of detecting strong sources that are not aligned with the galaxy catalog. For this case, the galactic likelihood ratio is uniformly taken to be unity.

To illustrate the capabilities of using the galaxy catalog in rejecting false GW+HEN coincidences, we calculated the probability that a random sky direction is within a given angular distance from at least one galaxy in the galaxy catalog. This is a simpler and less sensitive way of utilizing information on galaxy locations than used in the method (which includes the blue-luminosity weight), but it already shows the usefulness of this additional information in background rejection. The results are shown in Figure \ref{figure:galaxycatalogdemonstration}. The probability of the coincidence of a random sky direction and at least one galaxy is evaluated for angular distances ranging from 0.1 to 10 degrees and considering galaxies with four different horizon distances from the observer. For these curves, it is possible to estimate the probability that a background neutrino be falsely associated with a host galaxy. We see that for an horizon distance of 50 Mpc (which is larger than the 40 Mpc used in the present analysis), there is one chance in two to get a false association if the position uncertainty is of order of one degree. The efficiency of the galaxy catalog to discard background neutrino triggers is directly connected to this probability (since background neutrinos coming from directions where there is no galaxy can be discarded). This result indicates that, within the 40 Mpc distance horizon, most background GW+HEN events will be farther from any galaxy than the typical angular uncertainty of HEN direction reconstruction, demonstrating the benefit of using the galaxy catalog, even for this simple model.

\section{Joint GW+HEN Analysis}
\label{section:jointanalysis}

\begin{figure*}
\begin{center}
\resizebox{0.95\textwidth}{!}{\includegraphics{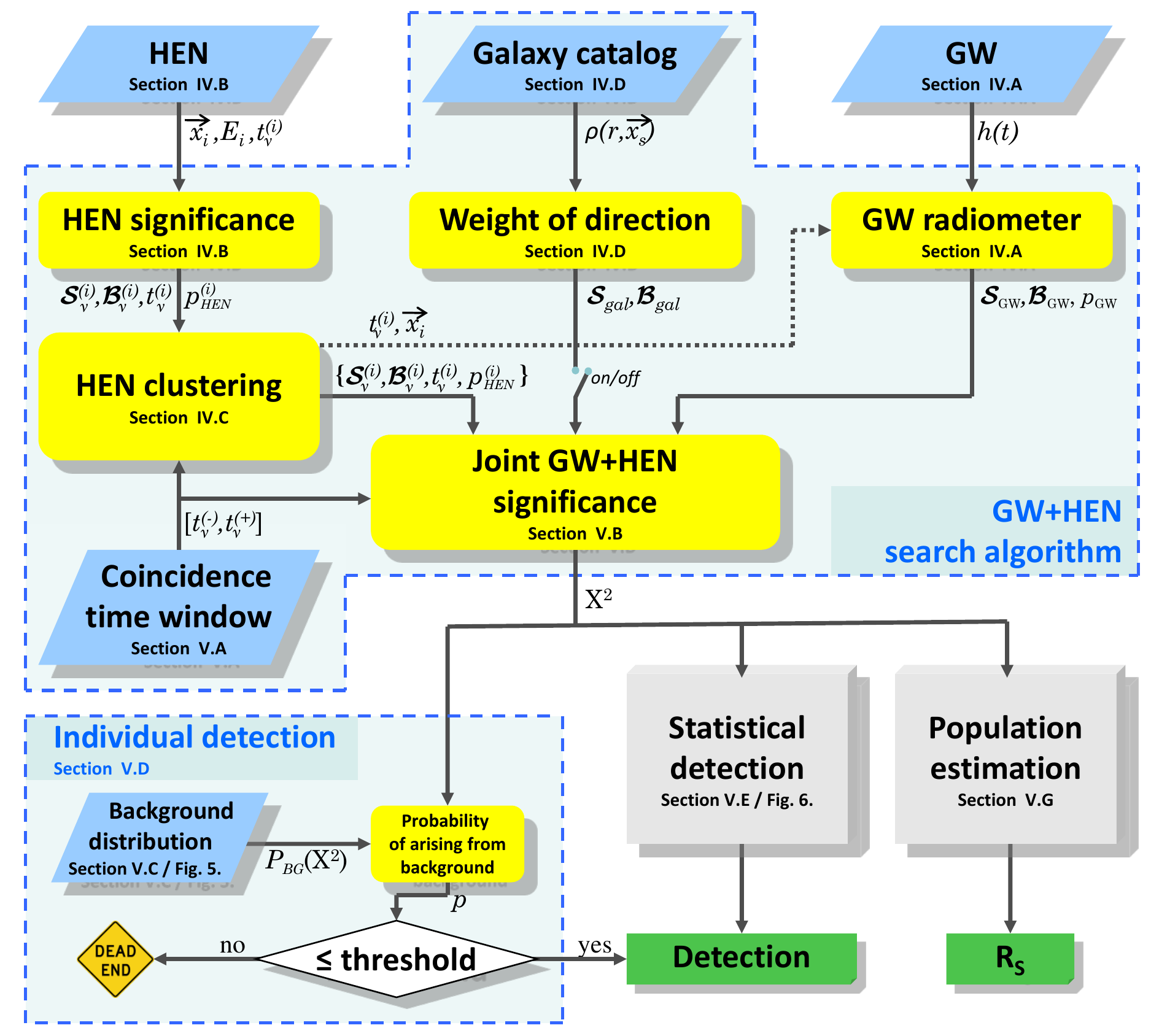}}
\end{center}
\caption{Flow diagram of the joint GW+HEN search algorithm. Steps include references to the sections and/or Figures in which they are described in details.} \label{figure:gwhenflow}
\end{figure*}

This section describes the joint analysis method for the search for GW+HEN signals. The joint analysis is described with a flow diagram in Fig. \ref{figure:gwhenflow}. For easier navigation the different steps in the flow diagram include references to the sections and figures where they are described in detail.

While the presented search example focuses on GWs and HENs, we note that it is straightforward to use the method with other messengers as well. The method also naturally incorporates externally triggered searches (see e.g. \cite{0004-637X-681-2-1419,2008PhRvL.101u1102A}), where at least one messenger confirms the presence of an astrophysical signal. A confirmed signal can define either or both the time window and source location (or point spread function) which restricts the parameter space of the multimessenger search a-priory. For such cases the interesting scientific question is whether additional information is present in other messengers, and if it is then what can one infer about the source by the total available information. We also note that the search can be analogously used for much lower energy neutrinos. For instance because the photomultipier tube (PMT) dark noise rate is particularly low in ice, the IceCube detector has sensitivity to sudden fluxes of MeV neutrinos which lead to collective rise in the PMT rates. Nearby supernovae up to 50 kpc are expected to be detected this way. While the MeV neutrino signal does not provide any directional information, it can be readily naturally incorporated in the present joint analysis by using its time of arrival and significance (i.e. flux). The lack of directional information can be taken into account as a uniform sky distribution. Further, similarly to the blue-luminosity-weighted galaxy distribution, a-priori source source distribution can be used for nearby sources as well, for example in the form of the matter distribution within the Milky Way. Galactic sources behind the center of the Milky Way can be especially interesting for multimessenger searches since they are difficult to observe electromagnetically.

\subsection{Coincidence Time Window}
\label{section:CoincidenceWindow}

The maximum time difference between the arrivals of the observed GW trigger and HEN events is one of the key parameters of the joint GW+HEN search algorithm \cite{Baret20111}. A too small time window might exclude some potential sources, while a too large time window would unnecessarily increase the false alarm rate and the computational cost.

Here, we adopt a conservative arrival time difference of $\pm500$~s derived for GRBs by Baret \emph{et al.} \cite{Baret20111}. Given a neutrino event, this allows for $\pm500$~s coincidence time window for a GW trigger. Multiple neutrino events and a GW trigger are considered temporally coincident if the greatest time difference between any two of these neutrinos, or any neutrino and the GW trigger, is less than 500~s.

We consider only one GW transient (trigger) per astrophysical GW+HEN source (we choose the GW trigger that gives the maximum joint significance; see below). Besides determining temporal coincidence, we apply no additional weight based on the arrival times of the HEN events and GW trigger (while the flux of neutrinos is probably time dependent, the uniform weight reflects our lack of information about this time dependence; see, e.g, \cite{2010APh....33..175B}).

\subsection{Joint GW+HEN Significance}

The joint significance combines the significances of the GW, HEN, and galaxy distribution components. For the directional distribution of these components, there exists a signal hypothesis, and therefore they are assigned a likelihood ratio. The other components of the GW and HEN events (e.g. energy) have no model-independent signal hypotheses, therefore they are assigned a p-value. These two types of information are combined into one joint significance measure.

\begin{figure*}
\begin{center}
\resizebox{0.95\textwidth}{!}{\includegraphics{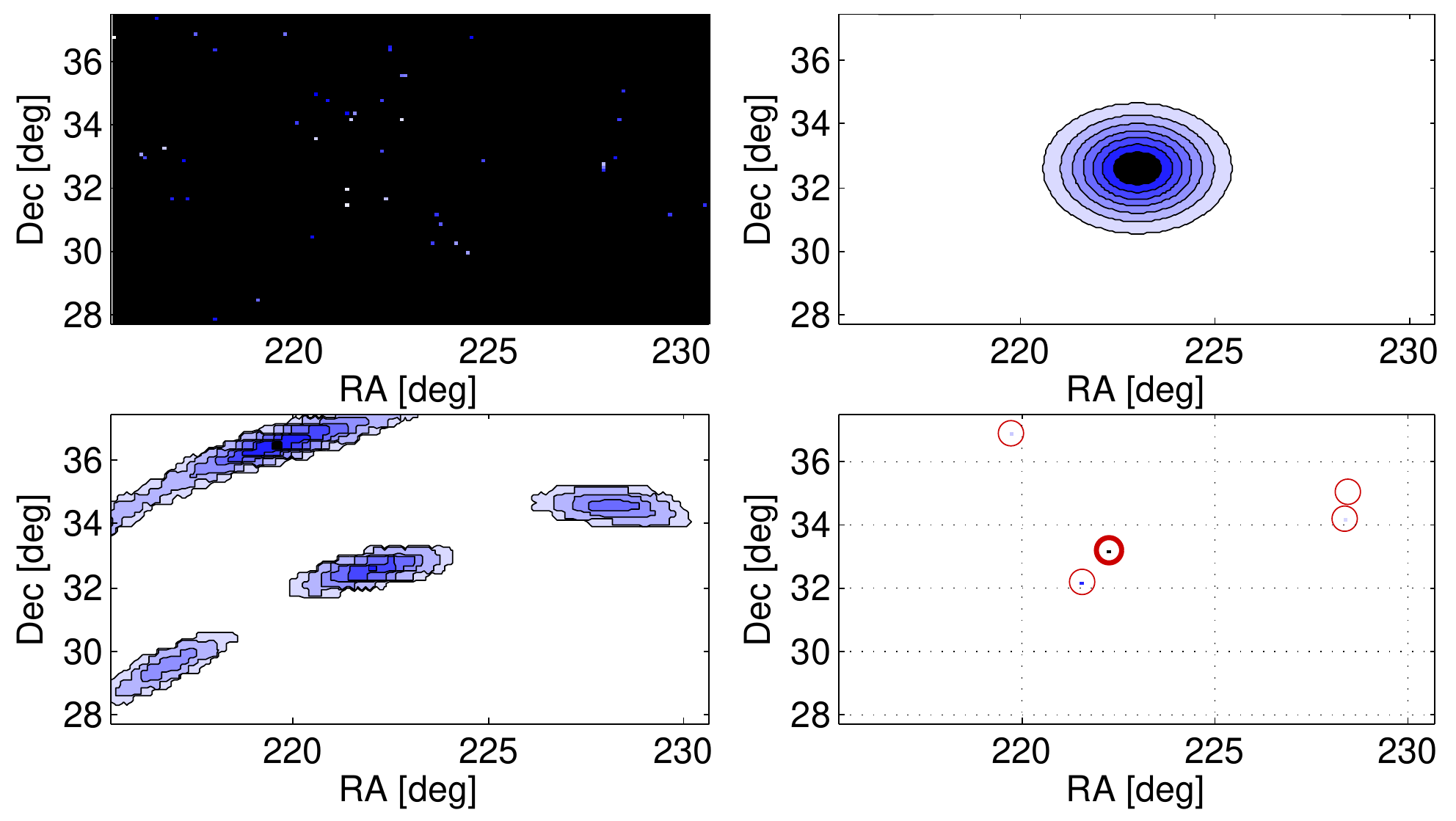}}
\end{center}
\caption{Example likelihood distributions as parts of the joint likelihood ratio: weighted galaxy distribution (upper left), HEN directional probability distribution function (PDF) (upper right), GW PDF (lower left) and  joint PDF (lower right). The joint PDF is the product of the other three PDFs. The scales are in arbitrary units, and the color-scale for the galaxy distribution is inverted. On the joint PDF plot, every galaxy for which the joint PDF is non-zero is circled for visibility. The reconstructed source direction with the maximum significance is circled with bold line.} \label{figure:lmaxexample}
\end{figure*}

First, we combine the likelihood ratios of the directional components to obtain a significance measure, i.e. p-value. The joint likelihood ratio $\mathcal{L}(\overrightarrow{x_s})$ is defined as
\begin{equation}
\mathcal{L}^{(i)}(\overrightarrow{x_s}) = \frac{\mathcal{S}_{\textsc{gw}}^{(i)}(\overrightarrow{x_s})\mathcal{S}_{gal}^{(i)}(\overrightarrow{x_s})\prod_{\{j\}}\mathcal{S}_{\nu}^{(j)}(\overrightarrow{x_s})}
{\mathcal{B}_{\textsc{gw}}^{(i)}\mathcal{B}_{gal}^{(i)}\prod_{\{j\}}\mathcal{B}_{\nu}^{(j)}},
\end{equation}
where $\{j\}$ is the set of neutrinos within GW+HEN trigger $i$. Note that the joint likelihood ratio, as it combines the directional distributions, is defined as a function of direction. Example directional distributions are shown in Fig. \ref{figure:lmaxexample}. Since we are mainly interested in the significance of the signal being of astrophysical origin, we treat the direction as a nuisance parameter and marginalize over it. Since the background likelihoods are uniform over the sky, the marginal likelihood ratio is
\begin{equation}
\mathcal{L}^{(i)} = \int \mathcal{L}^{(i)}(\overrightarrow{x_s}) \mbox{d}\overrightarrow{x_s}
\end{equation}
The background distribution of $\mathcal{L}$, $P_{\textsc{bg}}(\mathcal{L})$, can be obtained from time scrambled data (see section \ref{section:BackgroundDataGeneration}). Using this distribution, the p-value $p_{sky}$ of the directional part of the joint event can be calculated as
\begin{equation}
p_{sky}^{(i)}=\int_{\mathcal{L}^{(i)}}^{\infty}P_{\textsc{bg}}(\mathcal{L}')\mbox{d}\mathcal{L}'.
\end{equation}

We follow Fisher's method in combining the p-values into one joint test statistic:
\begin{equation}
X^{2}_{i} = -2\ln\left[ p_{sky}^{(i)}p_{\textsc{gw}}^{(i)}p_{cluster}(N)\prod_{\{j\}}p_{\textsc{hen}}^{(j)} \right]
\end{equation}
To ensure that possible correlations between the different p-values do not affect the outcome, we calculate the significance of $X^{2}$ by calculating its p-value from its background distribution $P_{\textsc{bg}}(X^{2})$ in time-scrambled data (see section \ref{section:BackgroundDataGeneration}):
\begin{equation}
p_{\textsc{gwhen}}^{(i)} = \int_{X^{2}_{i}}^{\infty}P_{\textsc{bg}}(X^{2'})\mbox{d}X^{2'}.
\label{totalsignificance}
\end{equation}

Note that, for a given joint event, we consider the single HEN or cluster of HENs with only one GW trigger at a time. If there are more GW triggers coincident with a given HEN trigger, we treat them as separate joint events (combined with the same neutrino).

To take into account possible sources outside of galaxies and in the Milky Way (or in galaxies not included in the used galaxy catalog), we perform an additional search without using the galaxy distribution. This is done simply by taking $\mathcal{F}_{gal}$ to be unity over the whole sky. This additional search without the galaxy catalog is represented by the on/off switch of the galaxy information in the flow diagram in Figure \ref{figure:gwhenflow}.

\subsection{Background Trigger Generation}
\label{section:BackgroundDataGeneration}

\begin{figure}
\begin{center}
\resizebox{0.5\textwidth}{!}{\includegraphics{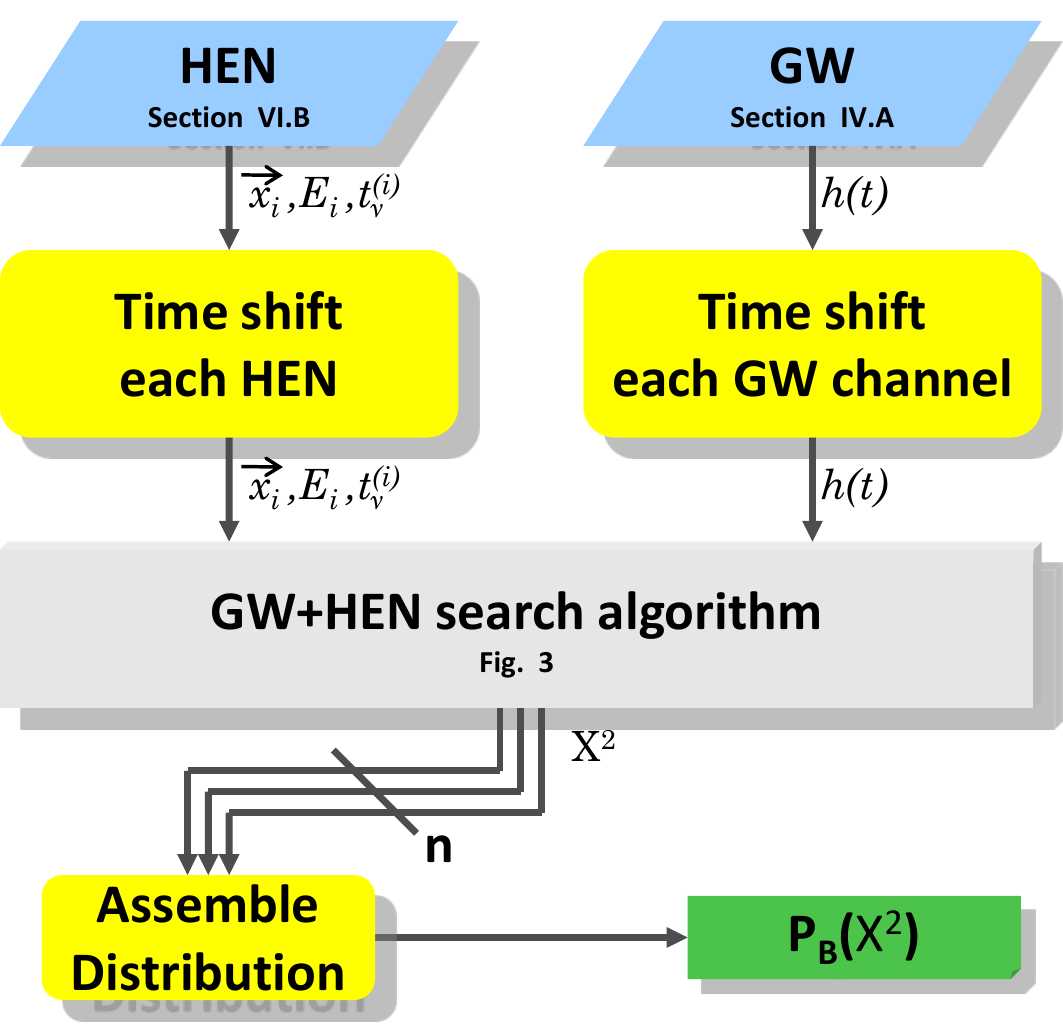}}
\end{center}
\caption{Flow diagram of the calculation of the background likelihood distribution $P_B(X^{2})$.} \label{figure:bglikelihoodflow}
\end{figure}

In order to calculate the significance of one or more joint signal candidates, we compare their test statistic $X^{2}$ to the test statistic distribution of background coincident triggers. The steps of the generation of the background distributions are described below, and are also shown in Figure \ref{figure:bglikelihoodflow}.

For a set of GW and HEN detectors, we apply time shifts for background event generation. For GW detectors the time shifts between any two detector data streams are selected to be much greater than the maximum possible time shift for an astrophysical signal. For neutrino detectors the time of arrivals are scrambled between neutrinos, keeping each event's local coordinates ($\theta$, $\phi$) and energy fixed during the scrambling.  This procedure reproduces fairly well the distribution of background neutrino parameters, and also preserves the geometric acceptances of the GW and HEN detectors which are fixed with respect to each other.

Using the background data described above, one can calculate the test-statistic distribution of background triggers similarly to how it is done for real data with no time shifts (see Eq. (\ref{totalsignificance})). 

\subsection{Individual Detection}
\label{section:SignificanceOfIndividualEventCandidates}

The loudest GW+HEN event in real data will be considered a \emph{joint detection} if its probability of arising from the background during a one year long measurement period is less than $2.87\times10^{-7}$ (one-sided 5$\sigma$).

We consider the joint GW+HEN to have \emph{discovery potential} for a given GW+HEN flux measured at Earth from an astrophysical source if such a signal has $50\%$ probability of resulting in a joint detection (as defined above).

We define the \emph{median upper limit} of the joint GW+HEN search to be the joint GW+HEN flux at Earth, in terms of $h_{rss}$ and $\langle n\rangle$, for which $X^{2}$ is greater, with $90\%$ confidence, than the median of the loudest events among sets of $N$ randomly generated joint background events.

To evaluate the statistical significance of a single GW+HEN event with $X^{2}$, we compare it to the background $X^{2}$ distribution $P_{\textsc{bg}}(X^{2})$ as described in section \ref{section:BackgroundDataGeneration}. The statistical significance of a $X^{2}$ value is given by its p-value (Eq. \ref{totalsignificance}).

\subsection{Statistical Detection of Multiple Sources}
\label{section:multisource}

\begin{figure}
\begin{center}
\resizebox{0.45\textwidth}{!}{\includegraphics{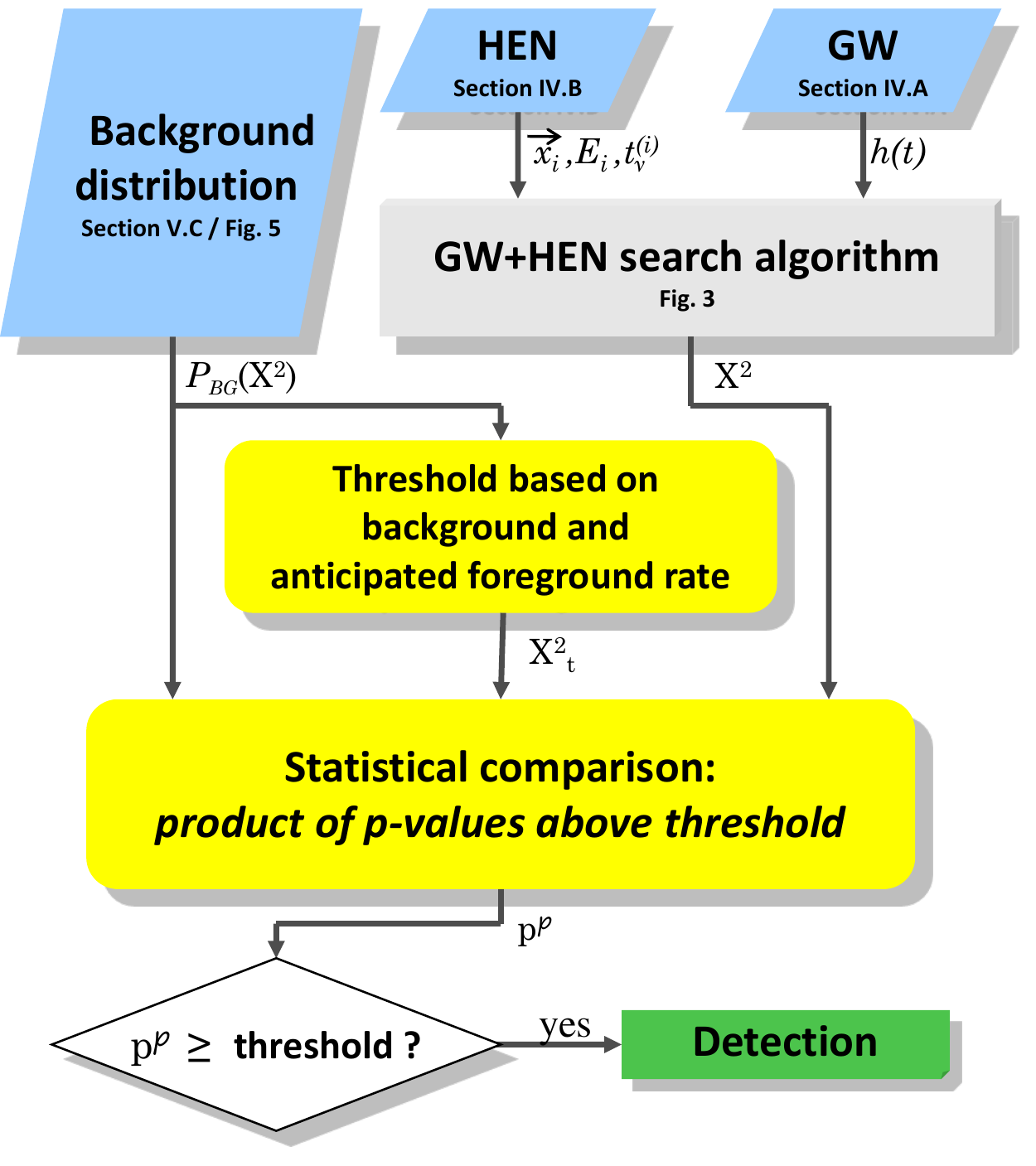}}
\end{center}
\caption{Flow diagram of the statistical detection of multiple GW+HEN sources.} \label{figure:statisticaldetection}
\end{figure}

Besides detecting a single astrophysical GW+HEN joint event, one can try to indicate the presence of multiple astrophysical joint events statistically. For such \emph{statistical detection}, we compare the distribution $P_\textsc{s}(X^{2})$ of the real data to the distribution $P_{\textsc{bg}}(X^{2})$ of joint background events. We use the realistic assumption that only a small fraction of the signal candidates can be due to actual astrophysical signals. The steps of statistical detection described below are also shown in Fig. \ref{figure:statisticaldetection}. Two alternative statistical tests can be found, e.g. in \cite{2008PhRvD..77f2004A}.

If only a small fraction of the joint events in real data are from astrophysical sources, one has the best chance of detecting the presence of real astrophysical signals by looking at the highest $X^{2}$ values. We therefore select a $X^{2}$ threshold above which the real-data and background distributions are compared. We denote this threshold by $X^{2}_{t}$. This threshold is chosen based on the background neutrino event rate, and the estimated astrophysical neutrino event rate within the distance in which the GW+HEN search is sensitive.

We compare the distributions of real and time-shifted data above $X^{2}_{t}$ in the following manner. Let $p_{t}$ be the p-value corresponding to $X^{2}_{t}$. We introduce the product $\mathfrak{p}$ for a set of p-values which are above threshold $p_{t}$. The value $\mathfrak{p}$ can be written as
\begin{equation}
\mathfrak{p}=\prod_{p^{(i)}_{\textsc{gwhen}}>p_{t}} p^{(i)}_{\textsc{gwhen}}
\end{equation}
where $p^{(i)}_{\textsc{gwhen}}$ is the p-value of measurement $i$ (see Eq. \ref{totalsignificance}). Similarly to the use of p-values for the single detection case, we calculate the probability $p^{\mathfrak{p}}$ that the measured $\mathfrak{p}$ from real data can arise from the background:
\begin{equation}
p^{\mathfrak{p}}=\int_{0}^{\mathfrak{p}}P_{\textsc{bg}}(\mathfrak{p}')\mbox{d}\mathfrak{p}',
\end{equation}
where $P_{\textsc{bg}}(\mathfrak{p})$ is the probability distribution of $\mathfrak{p}$ on time-shifted data (of identical duration as real data). The value $p^{\mathfrak{p}}$ is therefore the probability that the product of the p-values smaller than $p_{t}$ from real data arose from the background. We use $p^{\mathfrak{p}}$ to characterize the significance of statistical detection. We claim statistical detection if the probability that the real-data $p^{\mathfrak{p}}$ arose from the background during one year of measurement is less than $2.87\times10^{-7}$ (one sided 5$\sigma$).

\subsection{Simulation of Astrophysical Signals}
\label{section:SimulationOfAstrophysicalSignals}

We use simulated astrophysical signals to characterize the sensitivity of the GW+HEN search algorithm. The simulation is designed such that the results are scalable for different GW and HEN emission fluxes, therefore our results can be used to constrain the parameter space of GW and HEN emissions. We simulate standard GW+HEN sources with identical intrinsic GW+HEN emission (energy and spectrum). Upper limits for such standardized source populations are conservative estimates, taking the average emission, compared to taking a distribution of emission energies.

For the simulations we assume that zero or one HEN event is detected from each source. This is the likely situation for the part of the HEN parameter space that is not constrainable by a km$^3$ HEN detector alone.

The simulation of an astrophysical joint event consists of the following steps.
\begin{enumerate}[1.]
\item For a given direction, generate a simulated astrophysical HEN event coming from the source direction. We use Monte Carlo simulations to generate a random reconstructed energy and directional uncertainty for such a neutrino, using a source neutrino energy spectrum. We then generate a reconstructed source direction for the neutrino, drawn from a 2D Gaussian distribution centered around the real source direction, with standard deviation obtained from the Monte Carlo simulation.
\item Generate simulated astrophysical GW event coming from the source direction. We inject a GW signal with a given amplitude and waveform into real GW data into each GW detector used in the analysis, taking into account the direction of the source. The amplitude of the injected GW is chosen from a range that covers the amplitude region of interest for the joint search.
\end{enumerate}

\subsection{Population Estimation}
\label{section:populationestimation}

We define \emph{population upper limit} for the joint emitters of GWs and HENs as the lowest population which would produce -- with $>90\%$ probability, a joint event with higher significance than the loudest GW+HEN event in real data. 

We obtain population estimates by calculating the probability of detection rate for every galaxy. We adopt the following notation: for an astrophysical source in galaxy $i$, the galaxy has blue luminosity $L_B^{(i)}$ and is at a distance $r_i$. Further, let $L_{\textsc{b}}^{\mbox{\textsc{mw}}}$ be the blue luminosity of the Milky Way, $R$ the source rate [per year per MWE galaxy], $T_m$ the measurement duration, and $f_b$ the neutrino beaming factor of the source.
The calculation is the following.

Given an astrophysical source in galaxy $i$, the probability that at least one HEN will be detected from this source is \cite{PhysRevLett.107.251101}
\begin{equation}
p(n\geq1|r,n_{\textsc{hen}})=1-F_{poiss}\left(0|(10\mbox{ Mpc}/r)^2 n_{\textsc{hen}}\right),
\end{equation}
where $F_{poiss}$ is the Poisson cumulative distribution function, and $n$ is the number of detected neutrinos from the source. Therefore for galaxy $i$ the average number $\widehat{N}_i$ of sources with at least one detected neutrinos during the measurement will be
\begin{equation}
\widehat{N}_i(R,T_m) = p(n\geq1|r_i,n_{\textsc{hen}})\cdot R / f_b \cdot T\cdot L_{\textsc{b}}^{(i)}/L_{\textsc{b}}^{\mbox{\textsc{mw}}}.
\end{equation}
The population upper limit is obtained from $\widehat{N}_i(R,T_m)$ by requiring the total number of detected neutrinos within $D^{\textsc{gwhen}}$ to be 2.3 during a one year long measurement. This is done by summing $\widehat{N}_i(R,T_m)$ over all galaxies on the hemisphere in which the neutrino detector is sensitive. For IceCube this is $\delta_i \geq0$ where $\delta_i$ is the declination of galaxy $i$. The population upper limit will be
\begin{widetext}
\begin{equation}
R^{\textsc{ul}}(E_{\textsc{gw}}^{iso},n_{\textsc{hen}}) = \frac{2.3f_bL_{\textsc{b}}^{\mbox{\textsc{mw}}}}{T_m\sum_{\{r_i> D^{\textsc{gwhen}}, \delta_i \geq0\}}\epsilon_{\textsc{gw}}(r_i)p(n\geq1|r_i,n_{\textsc{hen}})L_{\textsc{b}}^{(i)}},
\end{equation}
\end{widetext}
where $\epsilon_{\textsc{gw}}(r_i)$ is the detection efficiency of the GW detector network at distance $r_i$ \cite{PhysRevD.81.102001}.

\subsection{Estimated Sensitivity}

In Section \ref{section:sciencereach} we estimated the science reach of the baseline multimessenger analysis following the calculations of Bartos et al. \cite{PhysRevLett.107.251101}. In estimating the science reach the single required parameter from the search algorithm was its exclusion distance $D_{50\%}^{\textsc{gwhen}}$. We approximated this parameter for the analysis by assuming that it will be comparable to the horizon distance of externally triggered GW searches. In the externally triggered GW search of Abbott \emph{et al.} \cite{0004-637X-715-2-1438}, the authors give the median horizon distance for single GW events coincident with electromagnetically observed GRBs. This horizon distance is related to the expected loudest GW background event from a given direction within a given time window of $\sim100$~s. The approximation of $D_{50\%}^{\textsc{gwhen}}$ for the baseline multimessenger search is reasonable if, given the number of HENs, the GW triggers' significance threshold is chosen such that the expected number of spatially and temporally coincident GW+HEN events is $\lesssim 1$. For one coincident event the remaining difference between the GW-GRB externally triggered search and the GW+HEN multimessenger search is mainly due to the greater directional uncertainty of neutrinos ($\sim1^\circ$) compared to the much better directional accuracy of electromagnetic GRB measurements. This difference, however, will not be significant, as the directional accuracy of GW measurements [O(10$^\circ$)] is much worse that of the HEN directional accuracy (see e.g. \cite{2011PhRvD..83j2001K,positionreconstructionfairhurst1367-2630-11-12-123006}). Further, the GW+HEN multimessenger analysis additionally takes into account the significance of HENs based on their reconstructed energy (see Section \ref{nu}) as well as the expected source distribution (see Section \ref{section:sourcedist}). Both of these pieces of additional information further increase the sensitivity of the search, making the comparison to results from externally triggered GW searches conservative.

To estimate the validity of the approximation that one will have $\lesssim 1$ joint event in a measurement without significant constraints on the rate of GW triggers, we take the example of the initial LIGO-Virgo detectors during their S5/VSR1 science run and the partially completed IceCube detector in its 22-string configuration. The LIGO-Virgo-IceCube network ran in coincidence from May 31, 2007 until Sept. 30, 2007 (123 days). IceCube-22, during its full run of 275.7 days, collected a final sample of 5114 neutrino candidate events \cite{1538-4357-701-1-L47}, of which $\sim 1000$ occur during the coincident livetime of the LIGO-Virgo-IceCube network. Considering a characteristic GW point spread function that spreads over 100 deg$^2$, the probability that temporally coincident GW and HEN triggers are also spatially coincident is $\sim0.5\%$. To reach on average $\approx1$ spatially and temporally coincident joint event during the measurement, we need on average 1 GW trigger for every 5 HEN triggers, corresponding to a GW trigger frequency of 17 day$^{-1}$. This limit is far from limiting the sensitivity of the search. For comparison, $\sim1$ GW trigger/day was used for electromagnetic follow-up observations for the latest LIGO-Virgo (S6/VSR2-3) science runs \cite{2011arXiv1109.3498T}.

Comparing the expected population upper limit of the GW+HEN multimessenger search to an all-sky GW search, the multimessenger search is expected to give stricter constraints on source population if its increased horizon distance compensates for HEN beaming (GWs are only weakly beamed). An approximate comparison is given by the ratio of the number of sources above the expected loudest-event threshold in each of the searches:
\begin{equation}
\frac{(D_{50\%}^{\textsc{gwhen}})^3/f_b}{(D_{50\%}^{\textsc{gw}})^3/f_{b,\textsc{gw}}},
\end{equation}
where $D_{50\%}^{\textsc{gw}}$ is the horizon distance of an all-sky GW search and $f_{b,\textsc{gw}}$ is the GW beaming factor. Taking $D_{50\%}^{\textsc{gwhen}} \approx 12$~Mpc from externally triggered GW searches \cite{0004-637X-715-2-1438}, $D_{50\%}^{\textsc{gw}} \approx 7.8$~Mpc from GW all-sky searches \cite{PhysRevLett.107.251101}, $f_b\approx14$ (an observational estimate for low-luminosity GRBs; \cite{Liang2007}) and $f_{b,\textsc{gw}}\approx 1.5$ (estimated value for inspirals or accretion-type GW emission; e.g. \cite{1538-4357-585-2-L89}), the ratio of detectable GW+HEN and GW events is $\approx0.4$, indicating that the number of sources excluded with the joint search is comparable to the number of those excluded with GW all-sky searches. Further, the sources probed by a joint search are mostly complementary to sources probed by an all-sky GW search. As the joint analysis is looking farther, it can potentially see sources missed by GW all-sky searches.

\section{Summary}
\label{section:conclusion}

We presented a baseline method for the joint search for GWs and significances of GW and HEN signals, including the point spread functions of event sky locations, as well as the blue-luminosity-weighted distribution of known galaxies.

We estimated the expected science reach of a joint GW+HEN search based on some of the existing GW and HEN emission models. The expected results indicate that the GW+HEN search with initial and particularly with advanced detectors will constrain the parameter space of some of the existing models.

To interpret the science reach of the expected GW+HEN population upper limits, we considered HEN rate expectations based on the Waxman-Bahcall emission model \cite{waxmanbachall}, as well as the emission model with mildly relativistic jets by Ando and Beacom \cite{chokedfromreverseshockPhysRevD.77.063007}.

The baseline GW+HEN analysis is expected to result in better upper limits than independent searches if the increase in exclusion distance compared to an independent GW search is greater than $\sim f_b^{1/3}$ (where $f_b$ is the HEN beaming factor; see also \cite{PhysRevLett.107.251101}). Further, the main advantage of a multimessenger search over individual searches is that of detection efficiency, as well as additional scientific information available from the source. Due to the non-Gaussianity of the GW data stream \cite{LIGO0034-4885-72-7-076901}, high-significance events have relatively high false alarm rate and therefore a coincident messenger can greatly increase our confidence in a detection. Additionally, joint detection would help better understand the underlying physics of the source. For instance a HEN detected prior to a GW from a GRB would indicate precursor activity in the GRB. The time delay of the earliest HEN after the GW signature of a collapsar may be indicative of jet propagation within the stellar envelope.

\section*{Acknowledgement}

The authors thank Kipp Cannon, Eric Howell, Eric Thrane for their useful comments. We are grateful for the generous support of Columbia University in the City of New York and the National Science Foundation under cooperative agreements PHY-0847182. The authors are grateful for support from the Swedish Research Council (VR) through the Oskar Klein Centre. We acknowledge the financial support of the National Science Foundation under grants PHY-0855044 and PHY-0855313 to the University of Florida, Gainesville, Florida. The authors thankthe authors thank the European Union FP7 (Marie Curie European Reintegration Grant NEUTEL-APC 224898). The authors acknowledge financial support from the French Agence Nationale de la Recherche (contract ANR-08-JCJC-0061-01) and from the EU FP7 (Marie Curie Reintegration Grant). The authors gratefully acknowledge the support of the United States National Science Foundation for the construction and operation of the LIGO Laboratory and the Science and Technology Facilities Council of the United Kingdom, the Max-Planck-Society, and the State of Niedersachsen/Germany for support of the construction and operation of the GEO600 detector. The authors also gratefully acknowledge the support of the research by these agencies and by the Australian Research Council, the Council of Scientific and Industrial Research of India, the Istituto Nazionale di Fisica Nucleare of Italy, the Spanish Ministerio de Educaci\'on y Ciencia, the Conselleria d'Economia, Hisenda i Innovaci\'o of the Govern de les Illes Balears, the Royal Society, the Scottish Funding Council, the Scottish Universities Physics Alliance, The National Aeronautics and Space Administration, the Carnegie Trust, the Leverhulme Trust, the David and Lucile Packard Foundation, the Research Corporation, the Alfred P. Sloan Foundation, and the French Centre National de la Recherche Scientifique.

\bibliographystyle{h-physrev}

\end{document}